\documentclass[epj]{webofc}
\usepackage[varg]{txfonts}   

\wocname{EPJ Web of Conferences}
\woctitle{Wigner 111 - Colourful \& Deep - Scientific Symposium}
\begin{document}
\title{Quantumness of discrete Hamiltonian cellular automata}
%
%

\author{Hans-Thomas Elze\inst{1}\fnsep\thanks{\email{elze@df.unipi.it}} 
}

\institute{Dipartimento di Fisica ``Enrico Fermi'', Universit\`a di Pisa,  
Largo Pontecorvo 3, I-56127 Pisa, Italia 
          }

\abstract{We summarize a recent study of discrete (integer-valued) Hamiltonian 
cellular automata 
(CA) showing that their dynamics can only be consistently defined, if it is linear 
in the same sense as unitary evolution described by the Schr\"odinger equation. 
This allows to construct an 
invertible map between such CA and continuous quantum mechanical  models, 
which incorporate a fundamental scale. 
Presently, we emphasize general aspects of these findings, the construction 
of admissible CA observables, and the existence of solutions of the modified dispersion 
relation for stationary states.}
\maketitle 

\section{Introduction}
The {\it Cellular Automaton Interpretation} of quantum mechanics has recently 
been laid out by G.  't\,Hooft \cite{tHooft2014}. The hope reflected in this far-reaching  
article, and in related works by others, is founded on the observation of quantum 
mechanical features arising in a large variety of deterministic ``mechanical'' models.  
While most of these models have been singular cases, i.e., which cannot easily be 
generalized to cover a realistic range of phenomena incorporating interactions, 
CA promise to provide the necessary versatility \cite{PRA2014,EmQM13}. 

The linearity of quantum mechanics (QM) is a fundamental feature most notably 
embodied in the Schr\"odinger equation. This linearity does not depend on the 
particular object  under study, provided it is sufficiently isolated from 
anything else. It is naturally reflected in the superposition principle and   
entails the ``quantum essentials'' interference and entanglement.    

The linearity of QM has been questioned repeatedly and nonlinear modifications 
have been proposed, in order to test experimentally the  
robustness of QM against such {\it nonlinear deformations}. 
This has been thoroughly discussed by T.F. Jordan presenting a stepwise proof 
`from within' QM that the theory has to be linear, given the 
additional {\it separability}   
assumption ``... that the system we are considering can be described as
part of a larger system without interaction with the rest of the larger 
system.'' \cite{Jordan}

Recently, we have considered a seemingly unrelated {\it discrete} 
dynamical theory, which appears to deviate drastically from quantum theory, 
at first sight.  
However, we have shown that the {\it deterministic} mechanics of the class of 
Hamiltonian CA can be 
related to QM in the presence of a fundamental time scale. This 
relation demonstrates that consistency of the action principle of the underlying discrete dynamics implies, in particular,  the linearity of both theories. 
This approach  may offer additional insight into  interference, entanglement, and  measurement processes in QM, in the limit when the 
discreteness scale is negligible.  

\section{CA Action Principle and observables}
The state of a classical cellular automaton (CA) with a denumerable set of degrees 
of freedom will be described by {\it integer-valued} ``coordinates''   
$x_n^\alpha ,\tau_n$ and ``conjugated momenta'' $p_n^\alpha ,\pi_n$, where 
$\alpha\in {\mathbf N_0}$ denote different degrees of freedom and $n\in {\mathbf Z}$  
different states. -- 
The {$x_n$ and $p_n$ might be higher dimensional vectors, while $\tau_n$ and 
${\cal P}_n$ are assumed one-dimensional. We separate the ``coordinate'' $\tau_n$ 
from the $x_n^\alpha$'s (correspondingly $\pi_n$ from the $p_n^\alpha$'s), 
since this degree of freedom represents the {\it dynamical time variable} here, 
discussed in \cite{PRA2014,EmQM13}, see also further references there. 
   
Finite differences, for all dynamical variables, are defined by: 
\begin{equation}\label{findiff}
\Delta f_n:=f_n-f_{n-1} 
\;\;. \end{equation} 
Furthermore, we define (with summation convention for Greek indices, 
$r^\alpha s^\alpha\equiv\sum_\alpha r^\alpha s^\alpha$)  
${\cal A}_n:=\Delta \tau_n (H_n+H_{n-1})+a_n\;$, 
$H_n:=\frac{1}{2}S_{\alpha\beta}(p_n^\alpha p_n^\beta+x_n^\alpha x_n^\beta  )
+A_{\alpha\beta}p_n^\alpha x_n^\beta + R_n\;$, 
$a_n:=c_n\pi_n\;$, 
where constants, $c_n$, and symmetric, $\hat S\equiv\{ S_{\alpha\beta}\}$,  and antisymmetric, $\hat A\equiv\{ A_{\alpha\beta}\}$, matrices are all integer-valued;  
$R_n$ stands for higher than second powers in $x_n^\alpha$ or  $p_n^\alpha$.  
The last definition  determines the 
behaviour of the variable $\tau_n$; a very simple 
choice suffices here, cf. below. 
 
Given these definitions, we introduce  the {\it integer-valued} CA {\it action}:  
\begin{equation}\label{action} 
{\cal S}:=
\sum_n[(p_n^\alpha +p_{n-1}^\alpha )\Delta x_n^\alpha 
+(\pi_n+\pi_{n-1})\Delta\tau_n
-{\cal A}_n]  
\;\;. \end{equation} 
For an alternative but equivalent form, which is particularly suited for the discussion 
of symmetry properties, see Ref.\,\cite{EmQM13}. -- Furthermore, let   
{\it integer-valued variations} $\delta f_n$ be applied to a polynomial $g$ in this way: 
\begin{equation}\label{variation} 
\delta_{f_n}g(f_n):=[g(f_n+\delta f_n)-g(f_n-\delta f_n)]/2\delta f_n 
\;\; , \end{equation} 
and $\delta_{f_n}g\equiv 0$, if $\delta f_n=0$. -- 
Then, CA dynamics is introduced by the following postulate.     
\vskip 0.25cm \noindent 
\underline{\it Action Principle}. \hskip 0.1cm   
The discrete evolution of a CA is determined by the stationarity of its 
action under arbitrary integer-valued variations of all 
dynamical variables, $\delta {\cal S}=0$.\,$\bullet$ \vskip 0.25cm  

Several features of this {\it Action Principle} are worth emphasizing: 
\\ \noindent  
{\bf i)} Variations  of terms that are 
{\it constant, linear, or quadratic (in dynamical variables)}  yield analogous results as  
infinitesimal variations of corresponding  
real-valued terms. \\ \noindent 
{\bf ii)} While infinitesimal variations do not conform with integer valuedness, 
there is {\it a priori} no restriction of integer variations, hence {\it arbitrary 
integer-valued variations} must be admitted. \\ \noindent 
{\bf iii)} However, for arbitrary  variations 
$\delta f_n$, the {\it remainder of higher powers} $R_n$ in $H_n$, which 
ultimately enters the action, has to vanish for consistency.  
Otherwise the number of equations of motion 
generated by variation of the action, generally, would exceed the number of 
variables. (However, a suitably chosen $R_0$ or a 
sufficient small number of such remainder terms can serve to encode  
the {\it initial conditions} for the CA evolution.)        
 
Employing the notation $\dot O_n:=O_{n+1}-O_{n-1}\;$,  
discrete analogues of Hamilton's equations are obtained by variation of 
the CA action ${\cal S}$ (keeping $R_n\equiv 0$):   
\begin{equation}\label{xdotCA} 
\dot x_n^\alpha\;=\;\dot\tau_n(S_{\alpha\beta}p_n^\beta +A_{\alpha\beta}
x_n^\beta ) 
\;\;,\;  
\;\;\dot p_n^\alpha \;=\;-\dot\tau_n(S_{\alpha\beta}x_n^\beta -A_{\alpha\beta}p_n^\beta ) 
\;\;, \end{equation}
\begin{equation}\label{taudotCA} 
\dot\tau_n\;=\;c_n 
\;\;,\;  
\;\;\dot\pi_n\;=\;\dot H_n 
\;\;, \end{equation} 
where all terms are integer-valued. Discreteness of the  {\it automaton time} $n$ is 
reflected by {\it finite difference equations} here.  Their appearance  
has motivated the name {\it Hamiltonian} CA.  

Further aspects of these equations, in particular the ensuing {\it symmetries  
and conservation laws}, have been discussed in Refs.\,\cite{PRA2014,EmQM13}. -- 
The equations are {\it time reversal invariant}. Most remarkably, they give rise to  conservation laws that are in {\it one-to-one correspondence} with those of the 
 Schr\"odinger equation for the {\it Hamilton operator} $\hat H$ given through the 
integer-valued 
symmetric and antisymmetric matrices (cf. above), $\hat S\equiv\{ S_{\alpha\beta}\}$  and 
$\hat A\equiv\{ A_{\alpha\beta}\}$, respectively: $\hat H:=\hat S+i\hat A\;$.   -- 
These observations are based on the fact that the Eqs.\,(\ref{xdotCA}) can be 
combined into: 
\begin{equation}\label{discrS} 
 \dot x_n^\alpha +i\dot p_n^\alpha =-i\dot\tau_n H_{\alpha\beta}
(x_n^\beta +ip_n^\beta ) 
\;\;, \end{equation} 
and its adjoint, employing the matrix elements of $\hat H$. 
This presents the {\it discrete analogue of Schr\"odinger's equation}, 
with $\psi_n^\alpha :=x_n^\alpha +ip_n^\alpha$ as the amplitude of the 
``$\alpha$-component'' of ``state vector'' $|\psi\rangle$ at ``time'' $n$

Presently, we would like to draw attention to another surprising parallel between the 
discrete and continuum models, CA and quantum mechanics, respectively. -- 
We may try to 
define a {\it ``Poisson bracket''} related to the dynamical variables of the CA, 
which are denoted collectively by $X_n,P_n$ and which represent the $x_n^\alpha, \tau_n$ and 
$p_n^\alpha ,\pi_n$, respectively: 
\begin{equation}\label{Poisson} 
\{ A,B\} :=\sum_n\left (\delta_{X_n}A\;\delta_{P_n}B-\delta_{X_n}B\;\delta_{P_n}A\right ) 
\;\;, \end{equation}   
employing the variational derivative defined in Eq.\,(\ref{variation}), 
since {\it ordinary derivatives are not available}; here 
$A$ and $B$ are polynomials depending on the dynamical variables.   

However, inspection shows that such polynomials $A$ and $B$ cannot be arbitrarily 
chosen, in order to have a consistent bracket which, besides showing 
{\it bilinearity}  and {\it antisymmetry}, also leads to {\it derivation-like product formula} 
and {\it Jacobi identity}, the defining properties of a Lie bracket operation.  --  
Namely, the problem arises that generally the result of the bracket operation might   
depend on the integer-valued variations $\delta f_n$, which enter through the definition 
of the 
variational derivative, Eq.\,(\ref{variation}). This would prohibit to form a closed 
algebra of polynomials. However, recalling observation 
{\bf i}) above, we restrict the polyomials to be 
{\it constant, linear, or quadratic (in dynamical variables)}. They form a closed 
algebra with respect to the  bracket operation, which becomes consistent in all respects. 

This simple result is remarkable for two reasons. -- First, the Hamilton operator $\hat H$ 
defines a quadratic form in terms of the $x_n^\alpha$ and $p_n^\alpha$, 
which can be compactly written as 
${\cal H}:=\sum_n\psi_n^{*\alpha}H_{\alpha\beta}\psi_n^\beta /2\;$. 
It corresponds to the expectation $\langle\psi |\hat H|\psi\rangle$ in quantum 
mechanics written in the particular representation developed by A. Heslot \cite{Heslot85}.   
This expectation belongs to the observables of a quantum mechanical object and should 
belong to the CA observables as well. In particular, since Eq.\,(\ref{discrS}) can be 
rewritten as  $\dot\psi_n^\alpha =\dot\tau_n\{ \psi_n^\alpha ,{\cal H}\}$. -- 
Second, restricting ourselves to {\it quadratic forms} in the dynamical variables as 
CA {\it observables} (eliminating trivial constant and linear forms that 
would yield inhomogeneous evolution equations), we arrive at a closed algebra of observables with respect to the Poisson bracket operation (\ref{Poisson}).  

We recall that all {\it quantum mechanical observables} are generated by 
Hermitean operators in this way  as quadratic forms \cite{Heslot85}.  -- 
Thus, insisting on the Hamiltonian structure of CA dynamics, including a suitably 
defined Poisson bracket, we are able to extend the close correspondence between 
CA and quantum mechanical systems to include the structure of the observables as 
well. 

\section{CA $\leftrightarrow$ QM map and modified dispersion relation}   
The correspondence that we discussed is not accidental and can 
be understood with the help of an {\it invertible map}  
between Hamiltonian CA and quantum mechanical objects that are characterized by 
a fundamental discreteness scale $l$. Implications for  
the conservation laws on both sides of the map were described in 
Refs.\,\cite{PRA2014,EmQM13}.  Here we reconsider  
the resulting dispersion relation, which might have  observable consequences. 

We employ Shannon's {\it Sampling Theorem} \cite{Jerri}:  
Consider square integrable {\it bandlimited functions} $f$, {\it i.e.}, which can be 
represented as $f(t)=(2\pi )^{-1}\int_{-\omega_{max}}^{\omega_{max}}\mbox{d}\omega\; 
\mbox{e}^{-i\omega t}\tilde f(\omega )$, with bandwidth $\omega_ {max}$. Given 
the set of amplitudes $\{ f(t_n)\}$ for the set  $\{ t_n\}$ of equidistantly spaced times  
(spacing $\pi /\omega_{max}$), the function $f$ is obtained for all $t$ 
by: 
\begin{equation}\label{samplingtheorem} 
f(t)=\sum_n f(t_n)\frac{\sin [\omega_{max}(t-t_n)]}{\omega_{max}(t-t_n)} 
\;\;. \end{equation} 
Since the CA ``time'' is given by an integer $n$, the discrete {\it physical time}  
is obtained by  multiplying with the scale $l$,  $t_n\equiv nl$, and the 
bandwidth by $\omega_{max}=\pi /l$. -- Next, we insert 
$\psi_n^\alpha :=x_n^\alpha +ip_n^\alpha$ in  Eq.\,(\ref{discrS})  and      
apply the {\it Sampling Theorem}, which maps this discrete time equation
invertibly to a  {\it continuous time equation}: 
\begin{equation}\label{modS}
2\sinh (l\partial_t)\psi^\alpha (t)
=\frac{1}{i}H_{\alpha\beta}\psi^\beta (t) 
\;, \end{equation} 
incorporating the simplest choice $\dot\tau_n\equiv 1$ . This is recognized as 
the {\it Schr\"odinger equation}, however, modified in important ways. -- 
The wave function $\psi^\alpha$ now  is bandlimited by $\omega_{max}$, which 
amounts to an 
{\it ultraviolet cut-off} of the energy $E$ of stationary states, 
$\psi_E(t):=\exp (-iEt)\tilde\psi$. Diagonalizing the self-adjoint  Hamiltonian,  
$\hat H\rightarrow\mbox{diag}(\epsilon_0,\epsilon_1,\dots )$,  
Eq.\,(\ref{modS}) yields the eigenvalue equation,
$\sin (E_\alpha l)=\epsilon_\alpha  /2=:\bar\epsilon_\alpha\;$, 
or, 
$E_\alpha =l^{-1}\arcsin (\bar\epsilon_\alpha )=
l^{-1}\bar\epsilon_\alpha [1+\bar\epsilon_\alpha^{\;2}/3!+\mbox{O}(\bar\epsilon_\alpha^{\;4})]\;$. Thus, we obtain a {\it modified dispersion relation}.  

Most importantly, discrete Hamiltonians do indeed exist  which have their 
spectrum bounded between -2 and 2, such that our eigenvalue equation has real solutions.   
A complete classification of such integer-valued {\it symmetric} matrices has recently 
been given \cite{McKeeSmyth}. This is a subject for future extension and physical 
interpretation, while all problems related to measurements in QM and their correlates 
in the Hamiltonian CA picture have still been left untouched. 

\vskip 0.1cm 
\noindent 
{\bf Acknowledgements:} I thank G. 't\,Hooft and L. Maccone for discussions and 
 correspondence, P. \'Ad\'am, T. B\'ir\'o, P. L\'evai and S. Varr\'o for inviting me to 
the inspiring ``Wigner 111 - Colourful \& Deep" symposium  (Budapest, November 2013), and L. Di\'osi for discussions and kind hospitality during this conference.

\end{document}